\documentclass[aps,prb,twocolumn,showpacs]{revtex4-1}
\pdfoutput=1
\usepackage{times}
\usepackage{epsfig}
\usepackage{amsmath}
\usepackage{amssymb}
\usepackage{wasysym}
\usepackage[colorlinks=true,linkcolor=blue, citecolor=blue]{hyperref}

\newcommand{\bs}{\boldsymbol}

\begin{document}

\title{Topological insulators from complex orbital order in transition-metal oxides heterostructures}
\author{Andreas R\"uegg and Gregory A. Fiete}
\affiliation{Department of Physics, The University of Texas at Austin, Austin, Texas 78712, USA}

\date{\today}

\begin{abstract}
Topological band insulators which are dynamically generated by electron-electron interactions have been theoretically proposed in two and three dimensional lattice models. We present evidence that the two-dimensional version can be stabilized in digital (111) heterostructures of transition-metal oxides as a result of purely local interactions. The topological phases are accompanied by spontaneous ordering of complex orbitals and we discuss their stability with respect to the Hund's rule coupling, Jahn-Teller interaction and inversion symmetry breaking terms. As main competitors we identify spin-nematic and magnetic phases.
\end{abstract}



\maketitle
\section{Introduction}
The search for new materials which realize a topological insulator (TI) phase\cite{Haldane:1988,Kane:2005b,Hasan:2011,Qi:2010} has dramatically increased, recently. Thereby, the main focus has been on compounds involving heavy elements with strong spin-orbit coupling.\cite{Hasan:2011,Qi:2010} On the other hand, it was pointed out that topological band properties can also arise from the spontaneous breaking of a symmetry in interacting systems where the spin-orbit coupling is negligible.\cite{Raghu:2008} In principle, this scenario suggests that TIs can be found in a much larger class of materials and several theoretical investigations support the existence of such interaction-driven TIs in two and three-dimensional interacting lattice models.\cite{Raghu:2008,Sun:2009,Zhang:2009,Weeks:2010,Wen:2010,Liu:2010b,Uebelacker:2011} From the experimental point-of-view, the situation is less satisfying and an experimental study of an interaction-driven TI is still lacking. In fact, the number of possible experimental systems is rather limited, and the most promising candidate so far is probably few-layer graphene.\cite{Nandkishore:2010,Zhang:2011}
In this work, we build on previous theoretical investigations and show that a two-dimensional interaction-driven TI phase may be stabilized from purely {\it local} interactions in multi-orbital models for transition-metal oxides. This result significantly extends the range of possible experimental systems. Using the conventional Hartree-Fock mean-field theory in combination with the theoretical analysis of the ${\bs k}\cdot {\bs p}$-model we find that the TI phase is accompanied by the {\it spontaneous ordering of complex orbitals}. 

\begin{figure}
\includegraphics[width=1\linewidth]{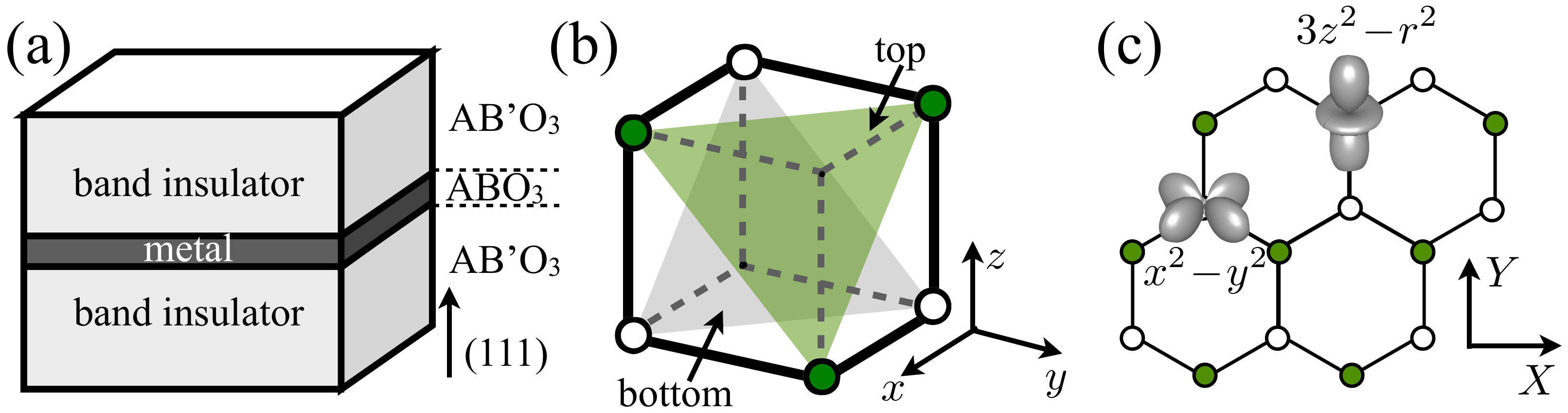}
\vspace{-0.1in}
\caption{(Color online.) (a) The digital oxide heterostructure considered in this article is grown in the (111) direction and of the form AB'O$_3$/ABO$_3$/AB'O$_3$. (b) The ``active" region consists of a (111) bilayer of the metallic ABO$_3$ perovskite. Shown are the locations of the transition-metal ions (B). (c) The bilayer system forms a honeycomb lattice when projected to the plane perpendicular to (111). We assume that the relevant orbital degrees of freedom are the $e_g$ orbitals of the transition-metal ions.}
\label{fig:setup}
\end{figure}

Our starting point is a system which belongs to the recently proposed class of digital oxide heterostructures\cite{Zubko:2011,Mannhart:2008} grown in the (111) direction\cite{Xiao:2011} and it is sketched in Fig.~\ref{fig:setup}. More precisely, we focus on the $d$-electrons of a (111) bilayer of the cubic transition-metal oxide ABO$_3$ [see Fig.~\ref{fig:setup}(b)] which is embedded in a band insulator AB'O$_3$. We assume that the ``active" compound (ABO$_3$) is metallic in bulk with a low spin $d^7$ configuration of the transition-metal (TM) ions, i.e. we assume filled $t_{2g}$ orbitals and one electron in the $e_g$ manifold. A possible choice of materials satisfying these requirements is a (111) bilayer of LaNiO$_3$ embedded in the band insulator LaAlO$_3$. 

The orbital degrees of freedom of the $e_g$ manifold are described by the two real orbitals $|a\rangle=|d_{z^2}\rangle$ and $|b\rangle=|d_{x^2-y^2}\rangle$, see Fig.~\ref{fig:setup}(c), which form a $T=1/2$-pseudo-spin $\vec{T}$. We will argue that topological phases can be stabilized by a spontaneous (and possibly spin dependent) ferro-orbital ordering of complex orbitals of the form
\begin{equation}
|d\pm id \rangle=\left(|d_{z^2}\rangle\pm i|d_{x^2-y^2}\rangle\right)/\sqrt{2}.
\label{eq:CO}
\end{equation}
These orbitals are eigenstates of $T^y$. In most cases, the complex orbitals Eq.~\eqref{eq:CO} are energetically disfavored because both super-exchange and lattice distortions prefer real orbitals in stoichiometric compounds.\cite{TMO:2004} However, the band structure of the considered (111) bilayer features a quadratic band crossing (QBC) point with a $d$-wave symmetry in orbital space which favors ordering of complex orbitals in a range of parameters at weak interactions.\cite{Sun:2009}

\section{Bilayer model}
As illustrated in Fig.~\ref{fig:setup}(b) and (c), the transition-metal ions of the (111) bilayer system form a honeycomb lattice. We study the following effective Hamiltonian for the $e_g$ electrons hopping on this honeycomb lattice
\begin{equation}
\mathcal{H}=\mathcal{H}_0+\mathcal{H}_{\rm int}+\mathcal{H}_{\rm perp}.
\label{eq:model}
\end{equation}
The band Hamiltonian $\mathcal{H}_0$ in the tight-binding approximation has been derived in Ref.~[\onlinecite{Xiao:2011}]. 
In momentum space it takes the form
\begin{equation}
\mathcal{H}_0=\sum_{{\bs k},\sigma}\vec{d}_{\sigma}^{\dag}({\bs k})H_0({\bs k})\vec{d}_{\sigma}({\bs k})
\end{equation}
where $\vec{d}_{\sigma}=(d_{1a\sigma}, d_{1b\sigma}, d_{2a\sigma}, d_{2b\sigma})^T$ is a vector of fermionic annihilation operators. Here, the bottom layer is labeled with the subscript $1$ and the top layer with the subscript $2$. The orbital labels are $a$ and $b$ and the spin is $\sigma$. The Bloch matrix $H_0({\bs k})$ is a $4\times 4$ matrix of the form
\begin{equation}
H_0({\bs k})=\begin{pmatrix}
0&0&\varepsilon_{a{\bs k}}&\varepsilon_{ab{\bs k}}\\
0&0&\varepsilon_{ab{\bs k}}&\varepsilon_{b{\bs k}}\\
\varepsilon_{a{\bs k}}^*&\varepsilon_{ab{\bs k}}^*&0&0\\
\varepsilon_{ab{\bs k}}^*&\varepsilon_{b{\bs k}}^*&0&0
\end{pmatrix}.
\label{eq:H0}
\end{equation}
Here, we kept only the dominant nearest-neighbor hopping $t$ and $\varepsilon_{a{\bs k}}=-t[1+\frac{1}{2}\cos(\frac{\sqrt{3}}{2}k_x)e^{-i\frac{3}{2}k_y}]$, $\varepsilon_{b{\bs k}}=-\frac{3t}{2}\cos(\frac{\sqrt{3}}{2}k_x)e^{-i\frac{3}{2}k_y}$ and $\varepsilon_{ab{\bs k}}=-i\frac{\sqrt{3}}{2}t\sin(\frac{\sqrt{3}}{2}k_x)e^{-i\frac{3}{2}k_y}$. [$k_x$ and $k_y$ directions refer to the $(X,Y)$-axes in Fig.~\ref{fig:setup}(c).] A more general form which also includes the second neighbor hopping is reproduced in the supplemental materials.\cite{TMOsupp,Slater:1954} Our main conclusions remain valid as long as $t$ is large compared to other tight-binding parameters. The non-interacting band structure of Eq.~\eqref{eq:H0} has an interesting and for the following discussion crucial feature: the Fermi surface at quarter filling consists of a single Fermi point ${\bs k}=0$ where two bands touch quadratically.\cite{Xiao:2011,LeeS:2011} This QBC point has a $d$-wave symmetry in orbital space and a six-fold rotation symmetry in ${\bs k}$-space which protects it from splitting into Dirac points.\cite{Sun:2009} We note here that the angular momentum of the $e_g$ manifold is quenched and spin-orbit coupling only enters as a higher-order process via coupling to the $t_{2g}$ orbitals\cite{Xiao:2011} which is assumed to be weak and neglected in the following. Furthermore, the linear coupling to the trigonal crystal field is also absent\cite{Xiao:2011} and the QBC point in the non-interacting band structure is a rather generic feature of the considered heterostructure.

The electron-electron interaction is accounted for by the local interaction between the $d$-electrons of the form
\begin{eqnarray}
\mathcal{H}_{\rm int}&&=\sum_{\bs r}\Big[U\sum_ {\alpha}n_{{\bs r}\alpha\uparrow}n_{{\bs r}\alpha\downarrow}+(U'-J)\sum_{\alpha>\beta,\sigma}n_{{\bs r}\alpha\sigma}n_{{\bs r}\beta\sigma}\nonumber\\
&&+U'\sum_{\alpha\neq \beta}n_{{\bs r}\alpha\uparrow}n_{{\bs r}\beta\downarrow}+J\sum_{\alpha\neq \beta}d_{{\bs r}\alpha\uparrow}^{\dag}d_{{\bs r}\beta\uparrow}d_{{\bs r}\beta\downarrow}^{\dag}d_{{\bs r}\alpha\downarrow}\nonumber\\
&&+I\sum_{\alpha\neq \beta}d_{{\bs r}\alpha\uparrow}^{\dag}d_{{\bs r}\beta\uparrow}d_{{\bs r}\alpha\downarrow}^{\dag}d_{{\bs r}\beta\downarrow}\Big].
\end{eqnarray}
The intra-orbital repulsion is denoted by $U$, the inter-orbital interaction by $U'$, $J$ parametrizes the Hund's rule coupling and $I$ the pair-hopping term. We employ the standard relations $U=U'+2J$ and $J=I$ valid for an isolated ion which leaves us with two independent interaction parameters $U$ and $J$.

The ideal electronic model is perturbed by $\mathcal{H}_{\rm perp}=\mathcal{H}_V+\mathcal{H}_{\rm JT}$. Here, $\mathcal{H}_V$ describes a sublattice potential which breaks the inversion symmetry between top and bottom layer
\begin{equation}
\mathcal{H}_V=\frac{V}{2}\sum_{{\bs k},\sigma\alpha}\left[d_{1\alpha\sigma}^{\dag}({\bs k})d_{1\alpha\sigma}({\bs k})-d_{2\alpha\sigma}^{\dag}({\bs k})d_{2\alpha\sigma}({\bs k})\right].
\label{eq:HV}
\end{equation}
This term is present if the bilayer system is capped by a different insulator than the one beneath it. Finally, $\mathcal{H}_{\rm JT}$ accounts for the cooperative Jahn-Teller effect which potentially drives a structural phase transition with distorted oxygen octahedra. The coupling of the electrons to the phonons of the oxygen displacements leads to an effective interaction between the electrons of neighboring transition-metal ions and we adapt the simple form\cite{TMO:2004,Nasu:2008}
\begin{equation}
\mathcal{H}_{\rm JT}=K\sum_{\langle i,j\rangle}\tau^l_i\tau_j^l.
\label{eq:HJT}
\end{equation}
The suffix $l=x,y,z$ denotes the direction of the bond between $i$ and $j$ and
$
\tau_i^l=\cos(\frac{2\pi n_l}{3})T_i^z-\sin(\frac{2\pi n_l}{3})T_i^z
$
with $(n_x,n_y,n_z)=(1,2,3)$. $K$ is positive and therefore favors a staggered orbital order of real orbitals.
\section{Phase diagram}
\begin{figure}
\includegraphics[width=1.0\linewidth]{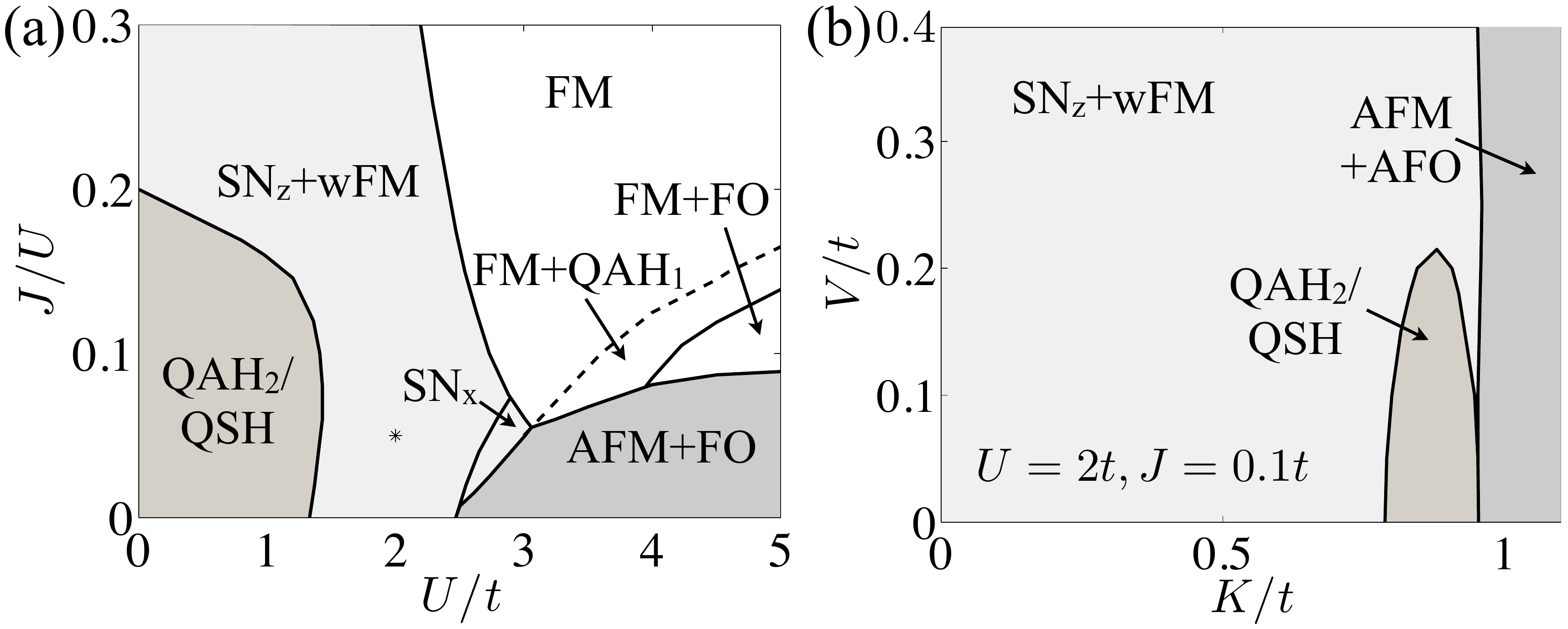}
\vspace{-0.1in}
\caption{(a) Zero temperature mean-field phase diagram of the ideal model as function of the repulsion $U/t$ and Hund coupling $J/U$. Topological phases are found in the small-$U$ limit (QAH$_2$/QSH) and within the ferromagnetic phase (FM+QAH$_1$). (b) Mean-field phase diagram as function of an inversion-symmetry breaking field $V/t$ and a Jahn-Teller interaction of strength $K/t$ for fixed interaction parameters [$\ast$ in (a)]. We find a topological phase between a spin nematic (SN$_z$) and an antiferromagnetic (AFM) phase. More details about the various phases are given in the main text.}
\label{fig:phasediagram}
\end{figure}

We first focus on the ideal electronic model and assume $\mathcal{H}_{\rm perp}=0$. The mean-field phase diagram obtained by 
solving the self-consistency equations numerically is shown in Fig.~\ref{fig:phasediagram}(a) as a function of the two dimensionless interaction parameters $U/t$ and $J/U$. We will understand the small-$U$ phases in this diagram qualitatively by analyzing the 
instabilities of the QBC in the next section. The strongly interacting limit is dominated by magnetic phases: If the Hund coupling $J/U$ is sufficiently small, we find an antiferromagnetic (AFM) phase which is accompanied by a ferro-orbital (FO) order. For larger ratios of $J/U$ we find (fully polarized) ferromagnetic (FM) order. In the absence of orbital order, the FM phase is gapless and has two Dirac nodes. Orbital order can open a gap in the FM phase. In particular, a quantum anomalous Hall state\cite{Haldane:1988} (QAH$_1$) with Chern number $n=\pm1$ is found if complex orbitals are involved.\cite{Yang:2011} The topological nature of this phase can be understood in analogy to the small-$U$ situation discussed below.

The weak to intermediate interaction regime is dominated by phases which are characteristic of the underlying QBC point.\cite{Uebelacker:2011} For small ratios $J/U$, we find an interaction-driven topological phase. The topological state either breaks the time-reversal symmetry  and has a finite Chern number $n=\pm2$ (QAH$_2$) or it preserves the time-reversal symmetry but breaks the spin-rotation symmetry realizing the quantum spin Hall (QSH) state.\cite{Kane:2005b} On the mean-field level, QAH and QSH phases are degenerate (in fact, for $I>J$ QAH and for $I<J$ QSH is favored).\cite{TMOsupp} 
The topological phase is surrounded by a {\it spin} nematic (SN) phase which also develops a weak FM order (wFM) for increasing $U/t$.  For $U/t\lesssim 0.9$ it is difficult to numerically resolve the energy difference between the topological and the spin nematic phases because both energies are exponentially small in $U/t$. The phase boundary shown in Fig.~\ref{fig:phasediagram}(a) for $U/t<0.9$ is an extrapolation to $J/U=0.2$ for $U/t\rightarrow 0$ which is the result obtained from the analysis of a reduced model in the next section. Finally, we note that charge nematic (CN) phases are absent for the considered parameters.\cite{TMOsupp}

We now briefly discuss some aspects of perturbing the ideal system with $\mathcal{H}_{\rm perp}=\mathcal{H}_V+\mathcal{H}_{\rm JT}$. In Fig.~\ref{fig:phasediagram}(b) we show the resulting phase diagram for fixed interaction parameters $U=2t$ and $J=0.1t$ which corresponds to the SN phase in the ideal model. For finite $K$ and $V$, the SN phase is accompanied by weak {\it ferri}-magnetic order. Interestingly, because the Jahn-Teller interaction favors {\it staggered} orbital order, it destabilizes the SN phase with {\it uniform} orbital order, allowing the topological phase to be energetically favored for some intermediate values of $K/t$ and small $V/t$. For even larger values of $K/t$, we find an AFM phase with staggered orbital order (AFO).

\begin{figure}
\includegraphics[width=1\linewidth]{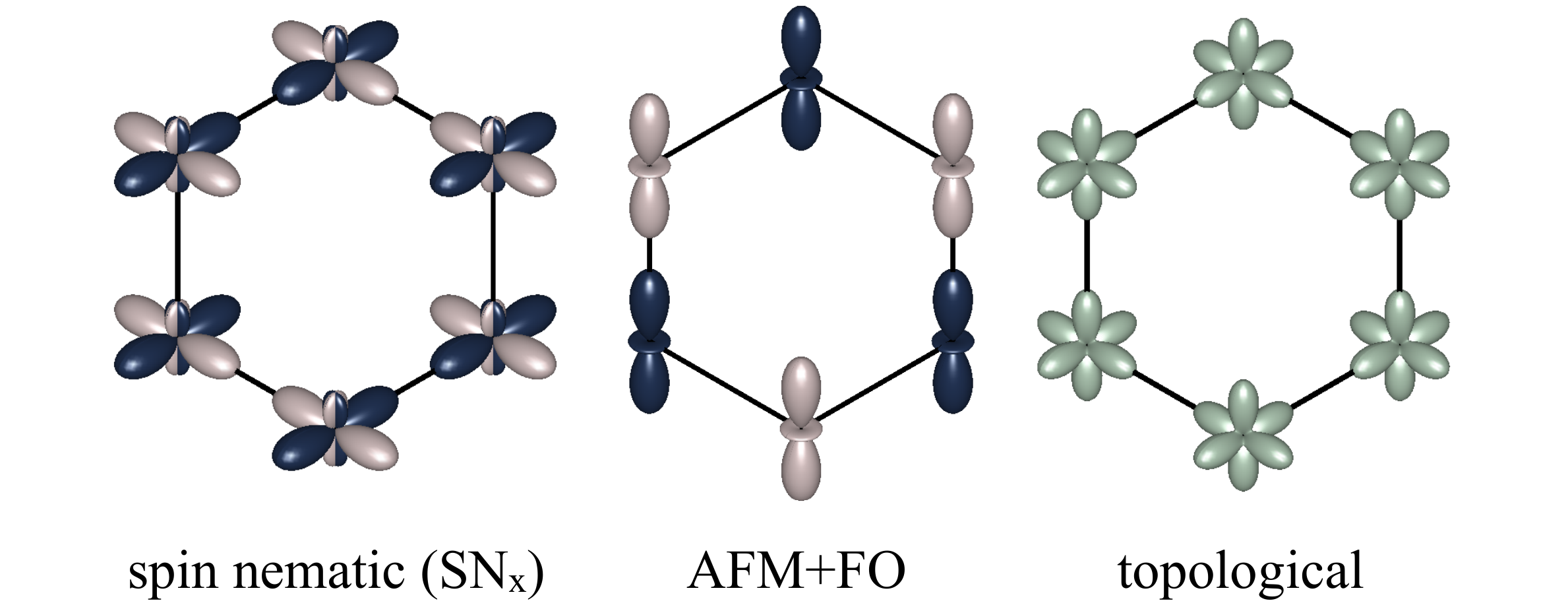}
\vspace{-0.1in}
\caption{Spin and charge densities in orbitally ordered phases of the bilayer system as seen from the (111) direction. The bright and dark orbitals in the spin nematic and AFM phase denote opposite majority spin densities and the orbital ordering also breaks the three-fold rotation symmetry. The relevant orbitals in the topological phase preserves the three-fold rotation symmetry.}
\label{fig:orbital_order}
\end{figure}

The mean-field analysis reveals various types of ordering of the orbital degrees of freedom. In Fig.~\ref{fig:orbital_order} we sketch the charge and spin density distribution for some representative examples. In the SN and the AFM phase the three-fold rotation symmetry of the lattice is broken and the orbital ordering involves real orbitals of the form
\begin{equation}
|\theta\rangle=\cos\frac{\theta}{2}|d_{z^2}\rangle-\sin\frac{\theta}{2}|d_{x^2-y^2}\rangle.
\label{eq:theta}
\end{equation}
The AFM phase orders in orbitals with $\theta_{\rm AFM}=0$ (or $\pm 2\pi/3$) which are eigenstates of $T_z$ (or the equivalent operators obtained by rotating $\vec{T}$ by $\pm 2\pi/3$ around the $y$ axis in orbital space). The spin nematic phase is either ordered along the $z$ axis (SN$_z$) or the $x$ axis (SN$_x$) in orbital space and electrons of a given spin are predominantly in one of the two orbital eigenstates. On the other hand, $\langle \vec{T}\rangle$ points along the $y$ direction for topological phases and the ordering involves complex orbitals of the form given in Eq.~\eqref{eq:CO}. As opposed to the real orbitals, the charge distribution associated with the complex orbital of the form Eq.~\eqref{eq:CO}  preserves the trigonal symmetry of the bilayer system. We note here that a finite spin-dependent ordering in the $y$-direction formally enters the mean-field Hamiltonian in the same way as the intrinsic spin-orbit coupling would.\cite{Xiao:2011} The relation between the topological band properties and the complex orbitals is further discussed in the next section.

\section{Reduced model for QBC point}

The competition among various weak coupling instabilities can be discussed in a reduced model which focuses only on the bands participating in the quadratic touching point and to momenta within a radius $\Lambda$ around the origin in ${\bs k}$-space (${\bs k}\cdot{\bs p}$-expansion). The effective model for the QBC point at the Fermi energy is found by expanding $H_0({\bs k})$ to order $k^2$ and eliminating the coupling to the higher bands in the same order by use of a canonical transformation. In polar coordinates the reduced Hamiltonian takes the form
\begin{equation}
\tilde{\mathcal{H}}_{0}=\!\!\sum_{\sigma}\!\int_{0}^{\Lambda}\!\!\frac{kd k}{2\pi}\!\!\int_0^{2\pi}\!\!\frac{d\phi}{2\pi}\, |k,\phi,\sigma\rangle\langle k,\phi,\sigma|\, \otimes \mathcal{H}_{\rm orb}(\phi,k).
\label{eq:H0red}
\end{equation}
$\mathcal{H}_{\rm orb}$ acts on the two-dimensional orbital space defined by the bonding orbitals of the bilayer system given by
\begin{equation*}
|\tilde{\alpha}\rangle=\frac{1}{\sqrt{2}}\left(|\alpha,1\rangle+|\alpha,2\rangle\right)
\end{equation*}
for ${\bs k}=0$ with $\alpha=a,b$. $\mathcal{H}_{\rm orb}$ has the standard form of a QBC point with $d$-wave symmetry\cite{Sun:2009}
\begin{equation}
\mathcal{H}_{\rm orb}(\phi,k)\!=\!k^2\left[t_{I} I\!+\!t_x\sin(2\phi)\tilde{T}_x\!+\!t_z\cos(2\phi)\tilde{T}_z\right].
\label{eq:H_orb}
\end{equation}
Here, $I$ denotes the identity and we have introduced the pseudo-spin operator $\vec{\tilde{T}}$ (with eigenvalues $\pm 1$) of the reduced orbital space $(|\tilde{a}\rangle,|\tilde{b}\rangle)$. The parameters in Eq.~\eqref{eq:H_orb} can be related to the hopping $t$ entering the full Bloch matrix Eq.~\eqref{eq:H0}:
$
t_I=t_z=-t_x=3t/16.
$
Diagonalizing Eq.~\eqref{eq:H0red} yields two quadratically dispersing bands with different effective masses which touch at ${\bs k}=0$: 
$
\epsilon_{1,2}({\bs k})=\mp k^2/(2m_{1,2}),
$
with
$
m_{1,2}=1/[2(|t_z|\mp t_I)].
$
(For $t_z=t_I>0$ the lower band is flat and $m_1\rightarrow\infty$.) The eigenfunctions of Eq.~\eqref{eq:H_orb} depend on the azimuth $\phi$ and have the simple form
\begin{equation*}
|1\rangle_{\phi}=\sin\phi|\tilde{a}\rangle+\cos\phi|\tilde{b}\rangle,\quad |2\rangle_{\phi}=\cos\phi|\tilde{a}\rangle-\sin\phi|\tilde{b}\rangle.
\end{equation*}
In other words, the orbital character changes {\it twice} when encircling the origin in ${\bs k}$-space. 

{\it Lifting the degeneracy: } 
The degeneracy of the spectrum at ${\bs k}=0$ is lifted by a uniform ``orbital field" $\vec{o}_{\sigma}$ which couples to the orbital pseudo-spin $\vec{\tilde{T}}$. A finite $y$-component $o_{y,\sigma}$ yields a $\phi$-independent coupling between the two bands. 
Adding a term $o_{y,\sigma}\tilde{T}_y$ to Eq.~\eqref{eq:H_orb} opens an energy gap throughout ${\bs k}$-space:
$
\varepsilon_{\pm,\sigma}({\bs k})=t_Ik^2\pm\sqrt{t_z^2k^4+o_{y,\sigma}^2}.
$
The resulting bands are topologically non-trivial and one finds a finite Chern number\cite{Thouless:1982,Qi:2006}
\begin{equation}
C_{\sigma}=\frac{1}{4\pi}\int\!\! d^2 k\,\, \hat{\bs n}\cdot\left(\frac{\partial\hat{\bs n}}{\partial k_x}\times\frac{\partial\hat{\bs n}}{\partial k_y}\right)=-{\rm sign}({o_{y,\sigma}}).
\label{eq:Csigma}
\end{equation}
Here, the limit $o_{y,\sigma}/(t_z\Lambda^2)\rightarrow0$ has been taken. The unit vector $\hat{\bs n}$ denotes the direction of the resulting field which couples to the orbital pseudo-spin $\vec{\tilde{T}}$:
\begin{equation*}
\hat{\bs n}=\left[-t_zk^2\sin(2\phi),o_{y,\sigma},t_z k^2\cos(2\phi)\right]^T/\sqrt{t_z^2 k^4+o_{y,\sigma}^2}.
\end{equation*}
$C_{\sigma}$ measures the winding number of the vector $\hat{\bs n}$ around the sphere. From Eq.~\eqref{eq:Csigma} we conclude that a finite orbital field in the $y$-direction results in a topological phase: If $o_{y,\uparrow}=o_{y,\downarrow}$, the two electrons occupy the same complex orbital realizing a QAH state with a total Chern number $n=\sum_{\sigma}C_{\sigma}=\pm 2$ and Hall conductivity $\sigma_{xy}=-n e^2/h$. If $o_{y,\uparrow}=-o_{y,\downarrow}$, time-reversal symmetry is preserved and $n=0$. However, this phase has a non-trivial $Z_2$-invariant and the QSH state is realized.

An orbital field which couples to $\tilde{T}_x$ and/or $\tilde{T}_z$ leads to a $\phi$-dependent coupling between the non-interacting bands.\cite{TMOsupp} As a result, the rotation symmetry of the spectrum in ${\bs k}$-space is broken and the QBC point splits into Dirac points with Berry phases $\pm\pi$.
Therefore, a orbital field in the $(x,z)$-plane corresponds to a nematic phase.\cite{Sun:2009,Wen:2010}

{\it Mean-field instabilities: }
We have analyzed the mean-field theory in the reduced space of the model Eq.~\eqref{eq:H0red} and have found that the order parameters of the nematic and topological phases enter through an orbital field which couples to $\vec{\tilde{T}}$. The CN and QAH order parameters enter through a spin-independent field while the SN and the QSH order parameters enter through a spin-dependent field with opposite values for $\uparrow$ and $\downarrow$ spins.\cite{TMOsupp} In either case, the linearized self-consistency equations can be solved and we find that the condensation energies depend exponentially on the interaction parameters. For the nematic phases $(\nu=$ SN, CN) we find
\begin{equation}
\Delta E_{\nu}\approx-2\gamma t_z^2\Lambda^4/u_{\nu}\exp\left(-8\pi t_z/u_{\nu}\right),
\label{eq:E_nu}
\end{equation}
where $\gamma\approx 10.8$ is a numerical factor.\cite{TMOsupp} The interaction parameters enter through the combinations $u_{\rm SN}=(U-J)/8$ and $u_{\rm CN}=(U-5J)/8$ and Eq.~\eqref{eq:E_nu} holds for $u_{\nu}>0$. Because $u_{\rm SN}>u_{\rm CN}$ for $J>0$ the spin nematic is favored over the charge nematic phase. 
For the topological phases we find the following condensation energy:
\begin{equation}
\Delta E_{\vartheta}\approx-2t_z^2\Lambda^4/u_{\vartheta}\exp\left(-4\pi t_z/u_{\vartheta}\right).
\label{eq:E_theta}
\end{equation}
The effective interactions for QAH and QSH are equal and given by $u_{\vartheta}=(U-3J)/8$. There is a factor of two different in the exponent of Eq.~\eqref{eq:E_nu} and Eq.~\eqref{eq:E_theta}. This difference can be traced back to the angular averaging in momentum space which for the nematic phases effectively reduces the interaction parameter in the exponent. The phase boundary between the topological and nematic phase is obtained by equating the condensation energies Eqs.~\eqref{eq:E_nu} and \eqref{eq:E_theta}. In the limit $U/t\rightarrow 0$, it suffices to compare the exponents which yields a critical ratio
$
\left(J/U\right)_c=1/5.
$
If $J/U<1/5$ the topological phase is preferred over the SN phase and if $J/U>1/5$ SN is preferred. For small but finite $U/t$ we find a monotonically decreasing phase boundary consistent with the numerical results of the full model presented in Fig.~\ref{fig:phasediagram}(a).
\section{Conclusion}
In summary, we have discussed a mechanism for spontaneous quantum Hall states in interacting multi-orbital models for a class of transition-metal oxide heterostructures. These topological phases are stabilized from purely local interactions and are accompanied by an orbital ordering of complex orbitals. In the weak coupling limit, the topological aspects can be understood qualitatively within a reduced model addressing the instabilities of the QBC point. Our results suggest that in the weakly interacting limit topological phases are most likely found for small Hund coupling in an inversion symmetric bilayer. Notably, we find that weak to intermediate Jahn-Teller interaction can help stabilizing a topological phase by suppressing its main competitor, the spin nematic phase. Furthermore, the mean-field calculations also suggest a topological phase at larger interaction strength within the strongly ferromagnetic regime.\cite{Yang:2011}

The (111) oxide heterostructures discussed in the present work offer a large freedom to design the electronic properties by suitable material combinations. Besides the two-dimensional spin-orbit\cite{Xiao:2011} or interaction-driven topological insulators one might also engineer more exotic topological phases such as the spin-charge separated QSH$^*$ phase\cite{Qi:2008,Ran:2008,Ruegg:2011b} or fractional quantum Hall states.\cite{Xiao:2011,WangF:2011,Neupert:2011a,Tang:2011,Sun:2011,Hu:2011,Sheng:2011,Neupert:2011,Qi:2011} An important future direction is to incorporate first-principle calculations to study the validity of the tight-binding approximation and to identify possible candidate materials.

\begin{acknowledgements}
We appreciate stimulating discussions with Alex Demkov and Allan MacDonald and acknowledge financial support through ARO grant W911NF-09-1-0527 and NSF grant DMR-0955778.
\end{acknowledgements}

%


\pagebreak
\onecolumngrid
\vspace{0.2in}
\begin{center}
{\bf \large Topological insulators from complex orbital order in transition-metal oxides heterostructures - Supplemental materials}
\end{center}
\vspace{0.1in}

\renewcommand{\thetable}{S\Roman{table}}
\renewcommand{\thefigure}{S\arabic{figure}}
\renewcommand{\thesubsection}{S\arabic{subsection}}
\renewcommand{\theequation}{S\arabic{equation}}

\setcounter{secnumdepth}{1}
\setcounter{equation}{0}
\setcounter{figure}{0}
\setcounter{section}{0}

\section{Band Hamiltonian}
\begin{figure}[h]
\includegraphics[width=0.6\linewidth]{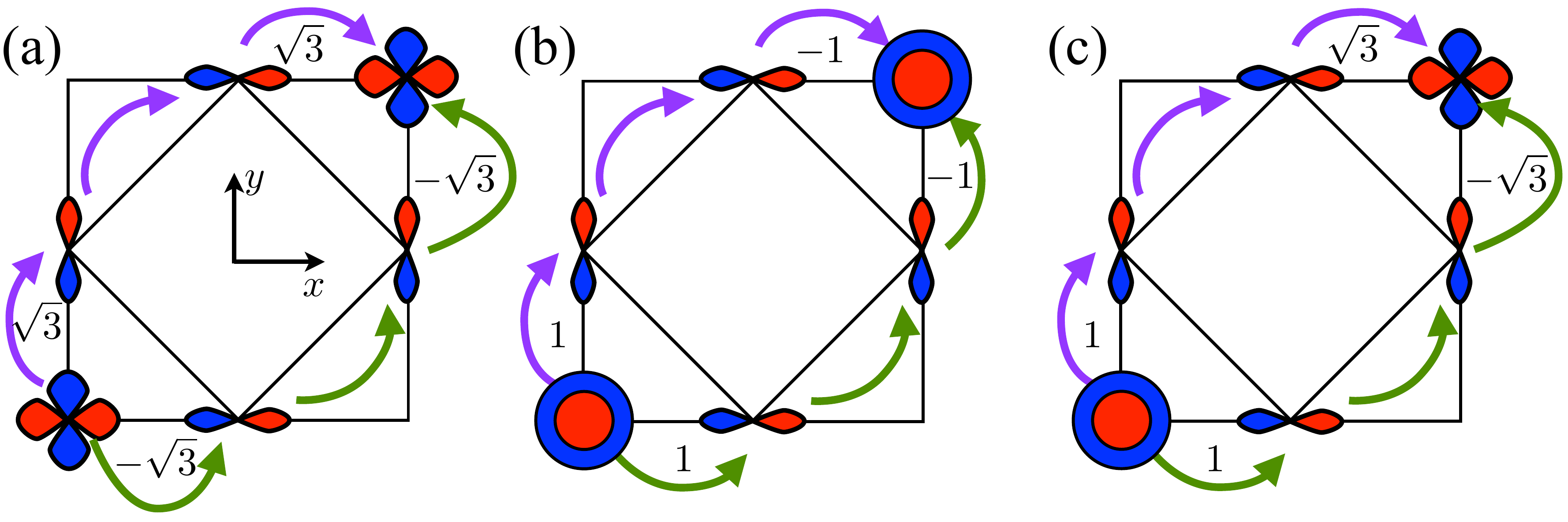}
\caption{Second-neighbor processes parametrized by the amplitude $t'$ shown for hopping in the $(x,y)$-plane. The numbers indicate the relative factors from the Slater-Koster energy integrals. (a) The diagonal second-neighbor hopping between $x^2-y^2$ orbitals and (b) between $3z^2-r^2$ orbitals. (c) The off-diagonal hopping vanishes because the two possible paths cancel.}
\label{fig:hoppings}
\end{figure}

Here, we reproduce the tight-binding Hamiltonian for the $e_g$ orbital degrees of freedom in the (111) bilayer system which also includs the direct overlap between the $d$-orbitals and higher-order processes which lead to the second neighbor hopping.\cite{Xiao:2011S}  The hopping matrix elements in real space are found from the Slater-Koster energy integrals.\cite{Slater:1954S} In momentum space, the hopping on the honeycomb lattice takes the form
\begin{equation}
H_0({\bs k})=\begin{pmatrix}
\tilde{\varepsilon}_{a{\bs k}}&\tilde{\varepsilon}_{ab{\bs k}}&\varepsilon_{a{\bs k}}&\varepsilon_{ab{\bs k}}\\
\tilde{\varepsilon}_{ab{\bs k}}&\tilde{\varepsilon}_{b{\bs k}}&\varepsilon_{ab{\bs k}}&\varepsilon_{b{\bs k}}\\
\varepsilon_{a{\bs k}}^*&\varepsilon_{ab{\bs k}}^*&\tilde{\varepsilon}_{a{\bs k}}&\tilde{\varepsilon}_{ab{\bs k}}\\
\varepsilon_{ab{\bs k}}^*&\varepsilon_{b{\bs k}}^*&\tilde{\varepsilon}_{ab{\bs k}}&\tilde{\varepsilon}_{b{\bs k}}
\end{pmatrix}.
\label{Seq:H0}
\end{equation}
The matrix elements are given by\cite{Xiao:2011S}
\begin{eqnarray*}
\varepsilon_{a{\bs k}}&=&-t-\frac{1}{2}(t+3t_{\delta})\cos(\frac{\sqrt{3}}{2}k_x)e^{-i\frac{3}{2}k_y},\\
\varepsilon_{b{\bs k}}&=&-t_{\delta}-\frac{1}{2}(3t+t_{\delta})\cos(\frac{\sqrt{3}}{2}k_x)e^{-i\frac{3}{2}k_y},\\
\varepsilon_{ab{\bs k}}&=&-i\frac{\sqrt{3}}{2}(t-t_{\delta})\sin(\frac{\sqrt{3}}{2}k_x)e^{-i\frac{3}{2}k_y},\\
\tilde{\varepsilon}_{a{\bs k}}&=&t'\left[4\cos(\frac{\sqrt{3}}{2}k_x)\cos(\frac{3}{2}k_y)-\cos(\sqrt{3}k_x)\right],\\
\tilde{\varepsilon}_{b{\bs k}}&=&3t'\cos(\sqrt{3}k_x),\\
\tilde{\varepsilon}_{ab{\bs k}}&=&2\sqrt{3}t'\sin(\frac{\sqrt{3}}{2}k_x)\sin(\frac{3}{2}k_y).
\end{eqnarray*}
Here, the unit of length $a$ is chosen as the nearest-neighbor distance on the honeycomb lattice and set to $a=1$. The first Brillouin zone is a hexagon with the two inequivalent $K$-points at $K_{\pm}=(\pm4\pi/(3\sqrt{3}),0)$.
The hopping parameter $t$ parametrizes the nearest-neighbor hopping due to the $\sigma$-hybridization with the oxygen $p$-states as well as the direct $\sigma$-bonding between neighboring $d$-orbitals. In terms of the Slater-Koster parameters,\cite{Slater:1954S} $t\sim (pd\sigma)^2/\Delta+(dd\sigma)$, where $\Delta$ is the energy difference between the oxygen $p$-states and the $e_g$-manifold of the transition-metal ion. The parameter $t_{\delta}\sim(dd\delta)$ parameterizes the remaining direct nearest-neighbor bonding. For the second-neighbor hopping $t'$, we have assumed that the most important processes are mediated by the bonding between nearest-neighbor oxygens as shown in Fig.~\ref{fig:hoppings}.

\section{Hartree-Fock approximation}
We have used the standard Hartree-Fock approximation for the local four fermion terms:
\begin{eqnarray}
d_{\alpha\sigma_1}^{\dag}d_{\beta\sigma_2}d_{\gamma\sigma_3}^{\dag}d_{\delta\sigma_4}&\rightarrow&\langle d_{\alpha\sigma_1}^{\dag}d_{\beta\sigma_2}\rangle d_{\gamma\sigma_3}^{\dag}d_{\delta\sigma_4}+d_{\alpha\sigma_1}^{\dag}d_{\beta\sigma_2}\langle d_{\gamma\sigma_3}^{\dag}d_{\delta\sigma_4}\rangle-\langle d_{\alpha\sigma_1}^{\dag}d_{\beta\sigma_2}\rangle \langle d_{\gamma\sigma_3}^{\dag}d_{\delta\sigma_4}\rangle\nonumber\\
&&- \langle d_{\alpha\sigma_1}^{\dag}d_{\delta\sigma_4}\rangle d_{\gamma\sigma_3}^{\dag}d_{\beta\sigma_2}-d_{\alpha\sigma_1}^{\dag}d_{\delta\sigma_4}\langle d_{\gamma\sigma_3}^{\dag}d_{\beta\sigma_2}+\langle d_{\alpha\sigma_1}^{\dag}d_{\delta\sigma_4}\rangle \langle d_{\gamma\sigma_3}^{\dag}d_{\beta\sigma_2}\rangle\rangle.
\end{eqnarray}
Additionally, we have assumed that possible magnetic order is co-linear and we have therefore only kept terms diagonal in spin-space
\begin{equation}
\langle d_{\alpha\sigma}^{\dag}d_{\beta\sigma'}\rangle=\delta_{\sigma\sigma'}\langle d_{\alpha\sigma}^{\dag}d_{\beta\sigma}\rangle.
\end{equation}
In other words, we only kept the $z$-components of the spin triplet-order parameters. Let us introduce the following operators which are diagonal in orbital space
\begin{eqnarray*}
\hat{n}&=&n_{a\uparrow}+n_{a\downarrow}+n_{b\uparrow}+n_{b\downarrow},\\
\hat{m}&=&n_{a\uparrow}-n_{a\downarrow}+n_{b\uparrow}-n_{b\downarrow},\\
\hat{p}&=&n_{a\uparrow}+n_{a\downarrow}-n_{b\uparrow}-n_{b\downarrow},\\
\hat{s}&=&n_{a\uparrow}-n_{a\downarrow}-n_{b\uparrow}+n_{b\downarrow}.
\end{eqnarray*}
Similarly, we introduce the following orbital off-diagonal operators
\begin{eqnarray*}
\hat{\chi}&=&d_{a\uparrow}^{\dag}d_{b\uparrow}+d_{a\downarrow}^{\dag}d_{b\downarrow}+{\rm h.c.},\\
\hat{\eta}&=&\frac{1}{i}\left(d_{a\uparrow}^{\dag}d_{b\uparrow}+d_{a\downarrow}^{\dag}d_{b\downarrow}-{\rm h.c.}\right),\\
\hat{\xi}&=&d_{a\uparrow}^{\dag}d_{b\uparrow}-d_{a\downarrow}^{\dag}d_{b\downarrow}+{\rm h.c.},\\
\hat{\lambda}&=&\frac{1}{i}\left(d_{a\uparrow}^{\dag}d_{b\uparrow}-d_{a\downarrow}^{\dag}d_{b\downarrow}-{\rm h.c.}\right).
\end{eqnarray*}
The expectation values of the above operators are denoted without hat ( $\hat{}$ ). 
\subsection{Local interaction}
With the above introduced notation and the assumption of collinear magnetic order the local interaction reduces to the following form
\begin{eqnarray}
\mathcal{H}_{\rm loc}&=&\frac{1}{4}(U+2U'-J)n\hat{n}-\frac{1}{8}(U+2U'-J)n^2\nonumber\\
&&-\frac{1}{4}(U+J)m\hat{m}+\frac{1}{8}(U+J)m^2\nonumber\\
&&-\frac{1}{4}(-U+2U'-J)p\hat{p}+\frac{1}{8}(-U+2U'-J)p^2\nonumber\\
&&-\frac{1}{4}(U-J)s\hat{s}+\frac{1}{8}(U-J)s^2\nonumber\\
&&-\frac{1}{4}(U'-2J-I)\chi\hat{\chi}+\frac{1}{8}(U'-2J-I)\chi^2\nonumber\\
&&-\frac{1}{4}(U'-2J+I)\eta\hat{\eta}+\frac{1}{8}(U'-2J+I)\eta^2\nonumber\\
&&-\frac{1}{4}(U'+I)\xi\hat{\xi}+\frac{1}{8}(U'+I)\xi^2\nonumber\\
&&-\frac{1}{4}(U'-I)\lambda\hat{\lambda}+\frac{1}{8}(U'-I)\lambda^2.
\label{eq:Hloc}
\end{eqnarray}
The QAH phase is characterized by $\eta\neq0$ and the QSH phase by $\lambda\neq 0$. Charge nematic phases are characterized by $(p,\chi)\neq(0,0)$ and spin nematic phases by $(s,\xi)\neq(0,0)$.
\subsection{Jahn-Teller interaction}
Following Ref.~[\onlinecite{TMO:2004S}], the cooperative Jahn-Teller effect gives rise to an effective interaction between the electrons of the form
\begin{equation}
H_{JT}=K\sum_{\langle i,j\rangle}\tau_{i}^{l}\tau_j^l
\end{equation}
with $K>0$. Here, $l$ denotes the direction of the nearest-neighbor bond $\langle i,j\rangle$ and we have introduced the following local pseudo-spin operators:
\begin{equation}
\tau_i^{l}=\cos\left(\frac{2\pi n_l}{3}\right)T_{iz}-\sin\left(\frac{2\pi n_l}{3}\right)T_{ix}
\end{equation}
with $(n_x,n_y,n_z)=(1,2,3)$. Explicitly, we obtain
\begin{equation}
\tau_i^x=-\frac{1}{2}T_{iz}-\frac{\sqrt{3}}{2}T_{ix},\quad
\tau_i^y=-\frac{1}{2}T_{iz}+\frac{\sqrt{3}}{2}T_{ix},\quad
\tau_i^z=T_{iz}.
\end{equation}
where the orbital pseudo-spin operators are defined as
\begin{equation}
\vec{T}_{i}=\frac{1}{2}\sum_{\sigma,a,b}d^{\dag}_{ia\sigma}\vec{\sigma}_{ab}d_{ib\sigma}
\end{equation}
with $\vec{\sigma}$ the Pauli matrices.
The eigenstates of these operators are the real orbitals of the form
\begin{equation}
|\theta\rangle=\cos\left(\frac{\theta}{2}\right)|d_{3z^2-r^2}\rangle+\sin\left(\frac{\theta}{2}\right)|d_{x^2-y^2}\rangle.
\end{equation}
In fact, the up pseudo-spin state for the operator $\tau^l_i$ corresponds to an occupied $d_{3l^2-r^2}$ orbital. In terms of the pseudo-spin $T$, the cooperative Jahn-Teller interaction is written as
\begin{eqnarray}
H_{JT}&=&\frac{K}{4}\sum_{i {\rm u. c.}}\left[T_{i}^zT_{i+x}^z+3T_{i}^xT_{i+x}^x+\sqrt{3}\left(T_{i}^zT_{i+x}^x+T_{i}^xT_{i+x}^z\right)\right]\nonumber\\
&&+\frac{K}{4}\sum_{i {\rm u. c.}}\left[T_{i}^zT_{i+y}^z+3T_{i}^xT_{i+y}^x-\sqrt{3}\left(T_{i}^zT_{i+y}^x+T_{i}^xT_{i+y}^z\right)\right]+K\sum_{i {\rm u. c.}}T_{i}^zT_{i+z}^z
\end{eqnarray}
where the summation runs over the unit cells of the honeycomb lattice. In the mean-field treatment, we consider the Hartree terms and assume a two-sublattice basis. Then, the expectation values are
\begin{equation}
\langle T_{iz}\rangle=\frac{1}{2}\begin{cases}
p_1, \mbox{ if } i\in \mbox{ bottom layer;}\\
p_2, \mbox{ if } i\in \mbox{ top layer.}
\end{cases}
\quad \langle T_{ix}\rangle=\frac{1}{2}
\begin{cases}
\chi_1, \mbox{ if } i\in \mbox{ bottom layer;}\\
\chi_2, \mbox{ if } i\in \mbox{ top layer.}
\end{cases}
\end{equation}
The Jahn-Teller interaction becomes
\begin{equation}
H_{\rm JT}=\frac{3K}{8}\sum_{i u.c.}\left[\hat{p}_{i,1}p_2+\hat{p}_{i,2}p_1+\hat{\chi}_{i,1}\chi_2+\hat{\chi}_{i,2}\chi_1\right]-\frac{3K}{8}\sum_{i u.c.}(p_1p_2+\chi_1\chi_2)
\end{equation}
where the summation is over the unit cells $i$. This form also shows that the Jahn-Teller interaction favors the staggered components of the real orbitals.

\section{Mean-field theory in the reduced space}
\subsection{Projection to the reduced space}
We first want to find a uniform transformation which brings the Bloch matrix Eq.~\eqref{Seq:H0} into block-diagonal form up to order $k^2$:
\begin{equation}
U^{\dag}H_0U=
\begin{pmatrix}
\tilde{H}_1&0\\
0&\tilde{H}_2
\end{pmatrix}+\mathcal{O}(k^3)
\end{equation}
Here, $\tilde{H}_1$ is the $2\times 2$ Bloch matrix of the lower QBC point and $\tilde{H}_2$ the Bloch matrix of the upper one.
The matrix $U$ can be found in two steps: $U=U_1U_2$. First, introduce the bonding and anti-bonding orbitals of the bilayer system by the transformation
\begin{equation}
U_1=\frac{1}{\sqrt{2}}\begin{pmatrix}
0&1&0&-1\\
1&0&-1&0\\
0&1&0&1\\
1&0&1&0
\end{pmatrix}
\end{equation}
which diagonalizes the Bloch matrix $H_0({\bs k})$ for ${\bs k}=0$ with eigenvalues $\mp 3/2(t+t_{\delta})+3t'$ and a energy difference $\Delta=3(t+t_{\delta})$ between upper and lower degeneracy points. However, the two sectors are coupled linearly in $k$:
\begin{equation}
U_1^{\dag}H_0U_1=
\begin{pmatrix}
H_1&T\\
T^{\dag}&H_2
\end{pmatrix},\quad T=T_xk_x+T_yk_y+\mathcal{O}(k^2).
\end{equation}
One can get rid of the linear coupling by introducing the second uniform transformation $U_2$:
\begin{equation}
U_2=e^{iS},\quad S=
\begin{pmatrix}
0&\tilde{S}\\
\tilde{S}^{\dag}&0
\end{pmatrix},\quad
\tilde{S}=-\frac{i}{\Delta}\left(k_xT_x+k_yT_y\right)
=\frac{1}{4(t+t_{\delta})}
\begin{pmatrix}
k_y(3t+t_{\delta})&k_x(-t+t_{\delta}\\
k_x(-t+t_{\delta})&k_y(t+3t_{\delta})
\end{pmatrix}.
\end{equation}
This then leads to the Bloch matrix $\tilde{H}_1$ of the reduced space which defines the model Eq.~\eqref{eq:H0red} with parameters
\begin{equation}
t_I=3\frac{t^2+6t_{\delta}t+t_{\delta}^2-12t'(t+t_{\delta})}{16(t+t_{\delta})},\quad t_z=-t_x=\frac{3(t-12t'-t_{\delta})}{16}.
\end{equation}
\subsection{Mean-field interaction in the reduced space}
Writing the local mean-field interaction Eq.~\eqref{eq:Hloc} in terms of the bonding and anti-bonding orbitals between top and bottom layer and keeping only the $k=0$ components acting on the reduced orbital space we find the following reduced mean-field interaction
\begin{equation}
\tilde{\mathcal{H}}_{\rm int}=-\sum_{\sigma}\!\int_{0}^{\Lambda}\!\!\frac{kd k}{2\pi}\!\!\int_0^{2\pi}\!\!\frac{d\phi}{2\pi}\, |k,\phi,\sigma\rangle\langle k,\phi,\sigma|\, \otimes \mathcal{H}_{\rm MF}(\sigma)+E_c.
\end{equation}
The mean fields act as (spin-dependent) orbital fields and $\mathcal{H}_{\rm MF}$ is given by
\begin{equation}
\mathcal{H}_{\rm MF}(\sigma)=\left(\vec{\tilde{p}}+\sigma\vec{\tilde{s}}\right)\cdot\vec{\tilde{T}}
\end{equation}
where the orbital pseudo-spins of the reduced subspace are defined by
\begin{equation}
\tilde{T}^x=|\tilde{a}\rangle\langle\tilde{b}|+|\tilde{b}\rangle\langle\tilde{a}|,\quad \tilde{T}^y=i\left(|\tilde{a}\rangle\langle\tilde{b}|-|\tilde{b}\rangle\langle\tilde{a}|\right),\quad \tilde{T}^z=|\tilde{a}\rangle\langle\tilde{a}|-|\tilde{b}\rangle\langle\tilde{b}|.
\end{equation}
The energy constant takes the form
\begin{equation}
E_c=\frac{1}{2u_{\rm CN}}(\tilde{p}_x^2+\tilde{p}_z^2)+\frac{1}{2u_{\rm SN}}(\tilde{s}_x^2+\tilde{s}_z^2)+\frac{1}{2u_{\vartheta}}(\tilde{\lambda}^2+\tilde{\eta}^2)
\end{equation}
and we have introduced $\tilde{\eta}\equiv\tilde{p}_y$ and $\tilde{\lambda}\equiv \tilde{s}_y$. The effective interactions are given by
\begin{equation}
u_{\vartheta}\equiv u_{p,y}=u_{s,y}=\frac{U-3J}{8},\quad u_{\rm CN}\equiv u_{p,x}=u_{p,z}=\frac{U-5J}{8},\quad u_{\rm SN}\equiv u_{s,x}=u_{s,z}=\frac{U-J}{8}.
\end{equation}
We note that the reduced mean-fields are proportional to the uniform components, i.e.,
\begin{equation}
\tilde{p}_l\sim p_{l,1}+p_{l,2},\quad \tilde{s}_l\sim s_{l,1}+s_{l,2}
\end{equation}
where $l=x,y,z$ and the subscript 1 and 2 refers to bottom and top layer, respectively. The staggered components enter only in higher order which is not considered here. Diagonalizing the full Hamiltonian $\tilde{\mathcal{H}}_0+\tilde{\mathcal{H}}_{\rm int}$ for finite order parameters yields the following eigenvalues
\begin{equation}
\epsilon_{\pm,\sigma}(k,\phi)=t_Ik^2\pm \sqrt{[t_z\sin(2\phi)k^2+\tilde{p}_x+\sigma \tilde{s}_x]^2+[t_z\cos(2\phi)k^2+\tilde{p}_z+\sigma \tilde{s}_z]^2+(\tilde{\eta}+\sigma\tilde{\lambda})^2}.
\end{equation}
\subsection{Self-consistency in the reduced space}
The self-consistency equations of the reduced model are given by
\begin{equation}
\frac{\tilde{p}_n}{u_{p,n}}=\sum_{\sigma}\!\int_{0}^{\Lambda}\!\!\frac{kd k}{2\pi}\!\!\int_0^{2\pi}\!\!\frac{d\phi}{2\pi}\,\langle -|\tilde{T}^n|-\rangle_{\phi,k,\sigma},\quad \frac{\tilde{s}_n}{u_{s,n}}=\sum_{\sigma}\!\int_{0}^{\Lambda}\!\!\frac{kd k}{2\pi}\!\!\int_0^{2\pi}\!\!\frac{d\phi}{2\pi}\,\langle -|\sigma\tilde{T}^n|-\rangle_{k,\phi,\sigma}.
\end{equation}
Here, $|-\rangle_{k,\phi,\sigma}$ denotes the orbital wave-function of a filled band which in general depends on $(k,\phi,\sigma)$.
\subsubsection{Topological phases}
The self-consistency equations for the topological phases ($\tilde{\eta}\neq 0$ or $\tilde{\lambda}\neq0$) assume the following form
\begin{equation}
\frac{1}{\pi}\int_0^{\Lambda}dk\frac{k}{\sqrt{k^4+k_{\vartheta}^4}}=\frac{t_z}{u_{\vartheta}}
\end{equation}
where $k_{\vartheta}^2=\tilde{\lambda}/t_z$ for QSH and $k_{\vartheta}^2=\tilde{\eta}/t_z$ for QAH. After performing the integration, the order parameter and the condensation energy are found:
\begin{equation}
k_{\vartheta}^2\approx 2\Lambda^2 \exp\left(-\frac{2\pi t_z}{u_{\vartheta}}\right) \quad\mbox{ and } \quad \Delta E_{\vartheta}=-\frac{k_{\vartheta}^4t_z^2}{2u_{\vartheta}}\approx -\frac{2\Lambda^4t_z^2}{u_{\vartheta}}\exp\left(-\frac{4\pi t_z}{u_{\vartheta}}\right).
\end{equation}
\subsubsection{Nematic phases}
Let us first introduce polar coordinates in the orbital $(x,z)$ plane:
\begin{eqnarray}
\tilde{p}_x=\tilde{p}\cos(\beta_p),&\quad& \tilde{p}_y=\tilde{p}\sin(\beta_s),\nonumber\\
\tilde{s}_x=\tilde{s}\cos(\beta_s),&\quad& \tilde{s}_y=\tilde{s}\sin(\beta_s).
\label{eq:polar}
\end{eqnarray}
It is then easy to show that the energy does not depend on the azimuth in Eqs.~\eqref{eq:polar} and the angles $\beta_{p,s}$ are undetermined in the reduced model. (Higher order terms entering the full Hamiltonian {\it do} select a definite direction, however). The self-consistency equations are more complex because the angular integration is non-trivial:
\begin{equation}
\frac{1}{\pi}\int_0^{\Lambda}\frac{k dk}{k_{\nu}^2}\frac{1}{k^2+k_{\nu}^2}\int_0^{2\pi}\frac{d\phi}{2\pi}\frac{k^2+k_{\nu}^2-2k^2\sin^2\phi}{\sqrt{1-m^2\sin^2(\phi)}}=\frac{t_z}{u_{\nu}}
\end{equation}
where $k_{\nu}=\tilde{p}/t_z$ for $\nu={\rm CN}$ (charge nematic) and $k_{\nu}=\tilde{s}/t_z$ for $\nu={\rm SN}$ (spin nematic) and we have introduced 
\begin{equation}
m=\frac{2kk_{\nu}}{k^2+k_{\nu}^2}.
\end{equation}
For $k_{\nu}\rightarrow 0$ the $k$ integration is dominated near $k=0$. We can therefore expand in small $m$:
\begin{equation}
\int_0^{2\pi}\frac{d\phi}{2\pi}\frac{k^2+k_{\nu}^2-2k^2\sin^2\phi}{\sqrt{1-m^2\sin^2(\phi)}}=k_{\nu}^2+\frac{k^2k_{\nu}^2}{k^2+k_{\nu}^2}-\frac{3k_{\nu}^2k^4}{2(k^2+k_{\nu}^2)^2}+\mbox{ regular terms},
\label{eq:exp}
\end{equation}
where the regular terms are of order one in the limit $k_{\nu}\rightarrow 0$ after $k$ integration. The first three terms, however, are logarithmically diverging for small $k_{\nu}$ and for $k_{\nu}\rightarrow 0$ we find
\begin{equation}
\frac{1}{4\pi}\log\frac{\Lambda^2}{k_{\nu}^2}+c\approx\frac{t_z}{u_{\nu}},
\end{equation}
where $c\approx0.150$ is a numerical constant which arises from the regular terms in the expansion Eq.~\eqref{eq:exp}.
It follows that the condensation energy for a nematic phase is given by
\begin{equation}
\Delta E_{\nu}=-\frac{k_{\nu}^4t_z^2}{2u_{\nu}}\approx -\frac{2\gamma\Lambda^4t_z^2}{u_{\nu}}\exp\left(-\frac{8\pi t_z}{u_{\nu}}\right)
\end{equation}
where $\gamma=\exp(8\pi c)/4\approx 10.8$ is a numerical factor.
\subsubsection{Boundary between topological phase and spin nematic}
The boundary between topological phase and spin nematic phase in the limit of small $U/t$ is given by equating the condensation energies of the reduced model, $\Delta E_{\vartheta}=\Delta E_{\rm SN}$, which can be written in the following form:
\begin{equation}
\gamma=\frac{1-y}{1-3y}\exp\left[\frac{a(1-5y)}{x(1-3y)(1-y)}\right],\quad x=\frac{U}{t},\quad y=\frac{J}{U},\quad a=\frac{32\pi t_z}{t}.
\end{equation}
The solution of the above equation defines the boundary $y(x)$. In the limit $x\rightarrow 0$ we find $y(0)=1/5$. For small but finite $x$ we find that $y(x)$ is monotonically decreasing, consistent with the numerical results obtained from the full model presented in Fig.~\ref{fig:phasediagram}(a) of the main part. 

\end{document}